\documentclass[conference]{IEEEtran}
\pdfoutput=1

\title{Fair Reward Distribution in Federated\\Byzantine Agreement Systems}

\author{
    \IEEEauthorblockN{
        Charmaine Ndolo\IEEEauthorrefmark{1},
        Martin Florian\IEEEauthorrefmark{2} and
        Florian Tschorsch\IEEEauthorrefmark{3}
    }
    \IEEEauthorblockA{
        \IEEEauthorrefmark{1}\IEEEauthorrefmark{2}\IEEEauthorrefmark{3}\textit{Humboldt-Universität
        zu Berlin},
        \IEEEauthorrefmark{2}\textit{Weizenbaum Institute}\\
    }
}

\usepackage{amsmath,amsthm,amssymb,amsfonts}
\newtheorem{theorem}{Theorem}
\newtheorem{definition}{Definition}

\usepackage[
    pdftitle={Fair Reward Distribution in Federated Byzantine Agreement
    Systems},
	pdfauthor={Charmaine Ndolo et al.}
]{hyperref}
\hypersetup{hidelinks,
  colorlinks=true,
  allcolors=black,
  pdfstartview=Fit,
  breaklinks=true}
\usepackage{cite}

\usepackage{enumitem}
\usepackage[babel=true]{csquotes}
\usepackage{mathtools}
\usepackage[nolist]{acronym}
\usepackage[colorinlistoftodos, textsize=tiny]{todonotes}
\usepackage[capitalise,nameinlink]{cleveref}
\usepackage{cuted}
\usepackage{pgfplots}
\usepackage{xcolor}
\definecolor{muted_indigo}{RGB}{51, 34, 136}
\definecolor{muted_cyan}{RGB}{136, 204, 238}
\definecolor{muted_teal}{RGB}{68, 170, 153}
\definecolor{muted_green}{RGB}{17, 119, 51}
\definecolor{muted_olive}{RGB}{153, 153, 51}
\definecolor{muted_sand}{RGB}{221, 204, 119}
\definecolor{muted_rose}{RGB}{204, 102, 119}
\definecolor{muted_wine}{RGB}{136, 34, 85}
\definecolor{muted_purple}{RGB}{170, 68, 153}

\pgfplotscreateplotcyclelist{COLOURS_LINES_MARKERS}{
    solid, every mark/.append style={solid, fill=muted_indigo, scale=1}, mark=*\\%
    dotted, every mark/.append style={solid, fill=muted_cyan, scale=1}, mark=square*\\%
    densely dotted, every mark/.append style={solid, fill=muted_teal, scale=1}, mark=otimes*\\%
    loosely dotted, every mark/.append style={solid, fill=muted_green,
    scale=1}, mark=triangle*\\%
    dashed, every mark/.append style={solid, fill=muted_olive, scale=1},mark=diamond*\\%
    loosely dashed, every mark/.append style={solid, fill=muted_sand, scale=1},mark=*\\%
    densely dashed, every mark/.append style={solid, fill=muted_rose, scale=1},mark=square*\\%
    dashdotted, every mark/.append style={solid, fill=muted_wine},mark=otimes*\\%
    dashdotdotted, every mark/.append style={solid, scale=1},mark=star\\%
    densely dashdotted,every mark/.append style={solid, fill=gray, scale=1},mark=diamond*\\%
}
\pgfplotsset{
    COLOUR_LINESTYLES_MARKERS_LIST/.style={
    cycle multiindex* list={
        muted_indigo, muted_cyan, muted_teal, muted_green, muted_olive,
        muted_sand, muted_sand, muted_rose, muted_wine, muted_purple\nextlist
        linestyles*\nextlist
        mark*\nextlist
    },
    },
}

\DeclareMathOperator{\V}{\mathbf{V}}
\DeclareMathOperator{\Q}{\mathbf{Q}}
\DeclareMathOperator{\I}{\mathcal{I}}
\DeclareMathOperator{\Minquorums}{\hat{\Quorums}}
\DeclareMathOperator{\players}{\mathit{N}}
\DeclareMathOperator{\charfunc}{\mathit{v}}
\DeclareMathOperator{\coalition}{\mathit{S}}
\DeclareMathOperator{\Pset}{2^N}
\DeclareMathOperator{\shapley}{\varphi}
\DeclareMathOperator{\sspi}{\phi}
\DeclareMathOperator{\W}{\mathit{W}}

\crefname{section}{Sec.}{Sec.}
\crefname{listing}{\lstlistingname}{\lstlistingname}
\Crefname{listing}{Listing}{Listings}
\crefname{definition}{Def.}{Def.}
\crefname{theorem}{Thm.}{Thm.}

\usetikzlibrary{positioning, plotmarks}
\usetikzlibrary{fit}
\usetikzlibrary{shapes.geometric}
\usetikzlibrary {backgrounds}

\DeclarePairedDelimiter\set{\{}{\}}

\newcommand{\dist}[1]{dist{#1}}
\newcommand{\Min}[1]{\hat{#1}}

\newcommand{\fbasdistlink}{\url{https://github.com/cndolo/fbas-reward-distributor/}}

\newcommand{\optithresh}[1]{\optithreshraw{\len{#1}}}
\newcommand{\optithreshraw}[1]{\ceil[\bigg]{\frac{2#1+1}{3}}}
\DeclarePairedDelimiter\len{\lvert}{\rvert}
\DeclarePairedDelimiter\ceil{\lceil}{\rceil}

\newcommand{\InnerQSets}{\mathcal{I}}
\newcommand{\Quorums}{\mathcal{U}}

\newcommand{\samples}{m}

\begin{acronym}
    \acro{NGT}{Non-Cooperative Game Theory} %
    \acro{CGT}{Cooperative Game Theory}
    \acro{FBAS}{Federated Byzantine Agreement System}
    \acro{MCP}{MobileCoin Consensus Protocol}
    \acro{MPE}{mean percentage error}
    \acro{MMPE}{mean mean percentage error}
    \acro{SCP}{Stellar Consensus Protocol}
    \acro{PoW}{Proof-of-Work}
    \acro{PoS}{Proof-of-Stake}
    \acro{P2P}{peer-to-peer}
\end{acronym}

\newcommand{\mycopyrightnotice}{
    \begingroup
    \renewcommand\thefootnote{}\footnote{
      \begin{minipage}{0.95\columnwidth}
          \copyright~2023 IEEE.
          Personal use of this material is permitted.
          Permission from IEEE must be obtained for all other uses, in any
          current or future media, including reprinting/republishing this
          material for advertising or promotional purposes, creating new
          collective works, for resale or redistribution to servers or lists, or
          reuse of any copyrighted component of this work in other works.
      \end{minipage}
  }
  \addtocounter{footnote}{-1}%
  \endgroup
}

\begin{document}
\maketitle
\mycopyrightnotice

\begin{abstract}
    \acfp{FBAS} offer a solution to consensus in permissionless systems by adapting the well-studied
    Byzantine agreement model to
    permissionless consensus.
    Unlike its counterparts in the context of permissionless consensus,
    the \acs{FBAS} system model does not offer validating nodes protocol-level
    incentives although they are entrusted with safeguarding and ensuring the
    functionality of the system.
    Multiple studies have reported on the small number of active validators in
    these systems leading to some concerns about their resilience.
    To this end, this paper studies how rewards can be distributed in \acp{FBAS}
    and presents a fair reward distribution function for \acp{FBAS}.
    The challenge is that, on the one hand, consensus in an \acs{FBAS} is found jointly between all
    nodes and, on the other hand, nodes do not all contribute equally to this process.
    We draw on game-theoretic methods to quantify these contributions bearing
    the overall health of the \acs{FBAS} in mind and present a fair reward
    distribution function which we evaluate based on a set of identified
    properties.
\end{abstract}

\section{Introduction}
\label{sec:intro}

\acfp{FBAS}~\cite{mazieres2015stellar} are a novel approach to consensus in
permissionless environments.
\acp{FBAS}
facilitate consensus in permissionless distributed systems
by allowing \emph{asymmetric} failure assumptions.
That is, each node is required to define a set of nodes
that it considers relevant during a consensus vote.
Provided that the sum of all such configurations
yields a set of desirable system properties,
consensus protocols can make use of the underlying \ac{FBAS} structure
to establish live and safe consensus.
The \acf{SCP}~\cite{mazieres2015stellar} builds on the \ac{FBAS} model and
steers the Stellar~\cite{stellar2019} and MobileCoin~\cite{mobilecoin2017}
payment networks.

\ac{FBAS} has been criticized for its lack of explicit incentives encouraging
nodes to participate in the network~\cite{kim2019stellar}.
In Bitcoin~\cite{nakamoto2008bitcoin} and the systems it inspired, fees and the
generation of new cryptocurrency units are used to reward a validator (miner)
for their services.
Recent studies on the Stellar and MobileCoin networks~\cite{kim2019stellar,
ndolo2021crawling} found that an overwhelming majority of the validators were
either operated by the networks' maintainers or organizations with economic
interests in the network.
As suggested in~\cite{ketsdever2019incentives}, the utility of such systems
serves as an implicit incentive for operators to join the network with
validating nodes.
However, in the case of the currently deployed FBAS-based networks, implicit
incentives are only sufficient to attract a fairly small and constant group of
entities.
We therefore maintain that explicit incentives are necessary not only for
network growth, but also in order to attract a more diverse set of operators
leading to decentralization.
In addition, validator rewards can also cover
the costs~\cite{florian2022sum, ndolo2021crawling} involved in maintaining a
validating node.

While numerous previous works deal with incentivization in blockchain-based
\acl{P2P} systems, to the best of our knowledge, incentivization in \acp{FBAS}
is yet to be addressed.
A great part of the literature on \acp{FBAS} covers the formal analysis of the
properties established by an \ac{FBAS} as well as the (efficient) computation
thereof~\cite{florian2022sum, gaul2019mathematical, kim2019stellar,
lachowski2019complexity}.

In this work, we present a reward distribution function for \aclp{FBAS}. Unlike
other permissionless consensus mechanisms, nodes in an \ac{FBAS} cooperate and
rely on each other to reach consensus
rendering a \enquote{winner-take-all} strategy inapplicable.
Hence, we investigate how rewards can be distributed fairly while taking into
account that some nodes provably contribute more to the functionality and
success of the \ac{FBAS}~\cite{florian2022sum,lachowski2019complexity}.
We define a set of requirements a reward distribution function for an \ac{FBAS}
should satisfy capturing the intuitive notions of equality, equity and
efficiency.
We conjecture that the sum of these requirements yields a fair distribution.
Our incentivization scheme is founded on determining each node's relative
\enquote{importance} using the Shapley value so as to define a distribution
function mapping the set of nodes to their rightful share of the reward.
To this end, we provide the following main contributions:
\begin{itemize}
  \item We map the \ac{FBAS} paradigm to the field of cooperative game theory and model an \ac{FBAS}
      as a cooperative game.
  \item We identify a set of requirements for a distribution function for
      \acp{FBAS} and define a consensus-protocol-agnostic reward distribution
      function based on the game-theoretic solution concept of the \emph{Shapley
      value}.
  \item We implement the reward distribution function, evaluate it formally and
      empirically using synthetic \acp{FBAS} against the set of requirements and
      provide optimizations for efficient computation.
\end{itemize}

The remainder is structured as follows.
In \cref{sec:prelim}, we introduce the \ac{FBAS} model
followed by a specification of the solution requirements.
Subsequently, we present a fair reward distribution in \cref{sec:design}.
In \cref{sec:evaluation}, we evaluate our approach.
We conclude the paper with an outline of related work in \cref{sec:related_work}
and summary of our results in \cref{sec:conclusion}.

\section{Federated Byzantine Agreement Systems}
\label{sec:prelim}
\label{sub:fbas}

\begin{figure}
  \begin{minipage}{0.5\columnwidth}
    \resizebox{\textwidth}{!}{
    \begin{tikzpicture}[node distance=4mm and 9mm]
      \node[draw,font=\footnotesize,thick,circle] (v0) at (0, 0) {0};
      \node[draw,font=\footnotesize,thick,circle,above left=of v0] (v1) {1};
      \node[draw,font=\footnotesize,thick,circle,below left=of v0] (v2) {2};
      \node[draw,font=\footnotesize,thick,circle,above right=of v0] (v3) {3};
      \node[draw,font=\footnotesize,thick,circle,below right=of v0] (v4) {4};
      \path [<->,thick,shorten >=2pt,shorten <=2pt] (v0) edge [draw] (v1);
      \path [<->,thick,shorten >=2pt,shorten <=2pt] (v0) edge [draw] (v2);
      \path [<->,thick,shorten >=2pt,shorten <=2pt] (v0) edge [draw] (v3);
      \path [<->,thick,shorten >=2pt,shorten <=2pt] (v0) edge [draw] (v4);
      \path [<->,thick,shorten >=2pt,shorten <=2pt] (v1) edge [draw] (v2);
      \path [<->,thick,shorten >=2pt,shorten <=2pt] (v3) edge [draw] (v4);
      \begin{scope}[on background layer]
        \node[circle,fill=gray,opacity=0.2,fit=(v0) (v1) (v2),inner sep=0pt] {};
        \node[circle,fill=gray,opacity=0.2,fit=(v0) (v3) (v4),inner sep=0pt] {};
        \node[ellipse,fill=gray,opacity=0.2,fit=(v1) (v4),inner sep=3pt] {};
      \end{scope}
    \end{tikzpicture}}
  \end{minipage}\hfill
  \begin{minipage}{0.4\columnwidth}
    \begingroup
    \addtolength{\jot}{-2pt}
    \scriptsize
    \begin{align*}
      \V &= \set{0,1,2,3,4}\\
      \Q(0) &= \set{\text{any } 3 \text{ out of } \set{0, 1, 2, 3, 4}}\\
      \Q(1) &= \set{\text{any } 3 \text{ out of } \set{0, 1, 2}}\\
      \Q(2) &= \set{\text{any } 3 \text{ out of } \set{0, 1, 2}}\\
      \Q(3) &= \set{\text{any } 3 \text{ out of } \set{0, 3, 4}}\\
      \Q(4) &= \set{\text{any } 3 \text{ out of } \set{0, 3, 4}}
    \end{align*}
    \endgroup
  \end{minipage}
    \caption{A simple FBAS ($\V, \Q$) illustrated as a directed graph in which
    there is an edge $(s,t)$ \emph{if} $t \in \Q(s)$.}
    \label{fig:fbas_example}
\end{figure}%

Mazières introduced \emph{\acfp{FBAS}}~\cite{mazieres2015stellar}
as an alternative approach to consensus in permissionless environments.
While classical Byzantine quorum systems enable consensus in systems with a
fixed number of nodes, the \ac{FBAS} model allows consensus in permissionless
environments by means of subjective \enquote{trust}.
Nodes are free to join (and leave) the system at free will, but every node in
the system is required to specify \emph{which} and \emph{how many} node failures
it will tolerate.
The underlying assumption is that nodes have some, but not necessarily global,
knowledge about the other nodes in the system enabling \emph{asymmetric} failure
tolerances.
Each node's failure tolerance is facilitated by off-chain decisions made by its
operator based on factors such as reputation and node longevity.
Configurations, however, may also be based on socio-technological factors such as
\enquote{trustworthiness}
as well as any other arbitrary factors considered by a node operator.

Compared to \acl{PoW}-based and \acl{PoS}-based systems, federated quorum systems take a different
approach to open membership in that they do not require participants to invest their own resources
as proof of legitimacy.
Hence, node operators are tasked with identifying legitimate, reliable
validators, e.g., via off-chain communication.
In the following, we outline core concepts about the \ac{FBAS} model
that are relevant to the remainder of this paper.

\subsection{Preliminaries}
\label{subsec:preliminaries}
An \ac{FBAS} is a set of nodes that runs a consensus protocol
such as the \acf{SCP}~\cite{mazieres2015stellar} which leverages the properties
of the established \ac{FBAS} and ensures the nodes reach agreement.
As nodes exchange messages in accordance with the governing consensus protocol,
each node chooses at least one set of nodes that must all accept a statement in
order for the selecting node to ratify the statement.
Such a set is known as a \emph{quorum slice}~\cite{mazieres2015stellar}.
Nodes are advised to define multiple quorum slices for the sake of fault
tolerance.
We define an \ac{FBAS} analogous to~\cite{mazieres2015stellar} as follows.

\begin{definition}[\acf{FBAS}~\cite{mazieres2015stellar}]\label{def:fbas}
    An \ac{FBAS} is a pair~$(\V,\Q)$ consisting of a set of nodes~$\V$
    and a quorum function~$\Q : \V \to 2^{2^{\V}} \setminus \emptyset$
    specifying quorum slices for each node.
    A node~$i$ belongs to all of its own quorum slices
    i.e., $\forall i \in \V, \forall q \in \Q(i), i \in q$.
\end{definition}

Stellar (and consequently MobileCoin) implements quorum slices as
\emph{quorum sets}~\cite{stellar2019, florian2022sum}.
Instead of specifying all possible sets that are able to convince a node, a node
defines a \emph{quorum set} comprising a set of nodes $U$, a threshold value
$t$, and inner quorum sets $\InnerQSets$.
A node $i$'s quorum set expresses the condition \enquote{at least $t$ nodes in
$U$} or \enquote{at least $t$ nodes in $U$ and $\InnerQSets$} must agree in
order for $i$ to assert a statement.
Implementations then generate all possible quorum slices based on a node's
quorum set.
The quorum set and quorum slice notations are equivalent~\cite{florian2022sum,
gaul2019mathematical}, and differ solely in their application:
the use of quorum slices has so far been limited to formalism
while the implementations hereof use quorum sets.
Throughout this paper, we will use an informal description
to denote a node's quorum slices
(values of~$\Q$) for comprehensibility and conciseness of notation.

As a running example,
\cref{fig:fbas_example} depicts an \ac{FBAS} ($\V, \Q$) with $\V = \set{0, 1, 2, 3, 4}$ and all
$\Q(i)$'s.
Informally, each of a node~$i$'s quorum slices~$\Q(i)$
describe the sets of nodes independently capable of
convincing $i$ to agree to a statement.
If $i$ expects that all members of one its quorum slices will externalize a value, %
so will~$i$.
A \emph{quorum} is a set of nodes that contains a quorum slice for each member of the set itself and
can, therefore, externalize new values by itself.

\begin{definition}[Quorum~\cite{mazieres2015stellar}]\label{def:quorum}
    A set of nodes $U \subseteq \V$ in \ac{FBAS} $(\V, \Q)$ is a quorum $\text{ iff } U \neq
    \emptyset$ and $U$ contains a slice for each member -- i.e. $\forall i \in U \; \exists q \in
    \Q(i) \colon q \subseteq U$.
    We denote the set of all quorums for $(\V, \Q)$ with $\Quorums \subseteq 2^{\V}$.
\end{definition}

The sets~$\Quorums = \set{\set{0, 1, 2}, \set{0, 3, 4}, \set{0, 1, 2, 3, 4}}$
form quorums for the \ac{FBAS} in \cref{fig:fbas_example}.
The set~$\set{0, 1 , 2}$, for example, satisfies node 0's, node 1's and node 2's
quorum slices meaning the three nodes are able to reach agreement among
themselves and externalize new values.
In what follows, and generally when reasoning about \acp{FBAS}, it is helpful to understand which
quorums exist.
We focus on the set of \emph{minimal quorums} $\Minquorums \subseteq \V$, i.e., a significantly
smaller set of quorums bearing sufficient information about the liveness of an \ac{FBAS}.

\begin{definition}[Minimal Quorums \cite{florian2022sum}]%
\label{def:minquorums}
    Let a \emph{minimal quorum} be a quorum $\Min{U} \subseteq \V$ for which there is no proper subset
    $U \subset \Min{U}$ that is also a quorum.
    We denote the set of all minimal quorums for $(\V, \Q)$ with $\Minquorums$.
\end{definition}

For our example \ac{FBAS} in \cref{fig:fbas_example}, the minimal quorums
are~$\Minquorums = \set{\set{0, 1, 2}, \set{0, 3, 4}}$.

Similar to sound distributed protocols, it is desirable that an FBAS-based
protocol can satisfy the \emph{liveness} and \emph{safety}
properties~\cite{alpern1985liveness}.
Liveness and safety in an \ac{FBAS} protocol are bounded by the structure of
the \ac{FBAS} itself~\cite{florian2022sum}.
That is, given an \ac{FBAS} $(\V, \Q)$,
\begin{itemize}
  \item Liveness is only achievable if the \ac{FBAS} enjoys
    \emph{quorum availability despite faulty nodes}, i.e.,
    there is a quorum of nodes $U \subseteq \V$ that contains only non-faulty nodes.
  \item Safety is only achievable if the \ac{FBAS} enjoys
    \emph{quorum intersection despite faulty nodes}, i.e.,
    $U_{1} \cap U_{2} \neq \emptyset$,
    where any two quorums intersect in at least one non-faulty node,
    independently of the behavior of faulty nodes.
\end{itemize}

Further discussions on the preconditions to safety can be found in~\cite{mazieres2015stellar}
and~\cite{florian2022sum}.
We focus on a node's contribution to liveness and later resort to the notion of
a \emph{top tier}~\cite{florian2022sum}, i.e., the set of nodes that is
exclusively relevant for determining the liveness properties of a given
\ac{FBAS} defined as follows.

\begin{definition}[Top Tier~\cite{florian2022sum}]\label{def:tt}
    The top tier of an FBAS $(\V, \Q)$ is the set of all nodes
    that are contained in one or more minimal quorums,
    i.e., if $\Minquorums \subseteq \Quorums \subseteq 2^{\V}$ is the set of all
    minimal quorums of the FBAS as defined in \cref{def:minquorums},
    $T=\bigcup{\Minquorums}$ is its top tier.
\end{definition}

\subsection{Problem Statement and System Model}%
\label{sub:problem_statement}

The networks Stellar and MobileCoin both register relatively modest numbers of
active validators~\cite{kim2019stellar, ndolo2021crawling}, i.e., 31 and 10
active validators respectively at the time of publication.
Kim et al.~\cite{kim2019stellar} presume
that the underwhelming participation is because of the lack of explicit
incentives and argue that an increase in the number of nodes leads to a more
secure network.

Since we concur with the analysis by Kim et al., our goal is to design a fair
reward distribution function for \acp{FBAS}.
This function should be independent of a particular consensus protocol used by
a specific \ac{FBAS} instance and should reward nodes relative to their
contribution to the system's health~(liveness).
We thus strive to define a reward distribution function that determines the
relative share of the reward a node~$i$ is entitled to receive per reward
period.
Any reward function requires an \ac{FBAS} ($\V, \Q$) as input.
Neither the actual value of the reward nor the source of the funds is subject of
this work, but pose interesting economic questions for future works.

We assume that the \ac{FBAS} $(\V, \Q)$ has quorum intersection as \acp{FBAS}
without quorum intersection cannot guarantee safety making forks and
double-spend attacks possible~\cite{mazieres2015stellar, florian2022sum}.
One reason for this assumption is that it is generally not possible to
distinguish an \ac{FBAS} that has accidentally lost safety from one that is
undergoing a legitimate split.
In general, an \ac{FBAS} without quorum intersection can be decomposed into
smaller \acp{FBAS} which enjoy quorum intersection independently.
This fact can be leveraged in practical deployments
when distributing rewards even if quorum intersection is lacking.

Furthermore, we assume a \enquote{global} view of the \ac{FBAS}
in that every node in $\V$ is in agreement about $(\V, \Q)$.
This simplification is necessary to guarantee a reward distribution that is
correct and verifiable for all nodes in the \ac{FBAS}.
It enables each node to compute its legitimate share of the reward locally.
In practice, it is plausible to assume that each node
has complete knowledge of the \ac{FBAS} given the availability and
instrumentality of such information when bootstrapping a node.
Snapshots of $(\V, \Q)$ could also be included in new blocks,
which are in turn agreed upon using a consensus protocol
such as \ac{SCP}.

\subsection{Requirements for Reward Distribution Functions}%
\label{subsec:requirements}

In the following,
we define a set of properties a reward distribution function
for nodes of an \ac{FBAS} should ideally fulfill.
This set of properties describes the basic characteristics of a sound reward
distribution scheme.
Let $i, j \in \V$ and $\dist(i) \colon \V \mapsto [0, 1]$ be a reward distribution function for the
\ac{FBAS} $(\V, \Q)$.
We classify $\dist(i)$ as a fair distribution function if it meets the following criteria:
\begin{enumerate}
    \item \label{req:eq} \emph{Symmetry}: Let ($\V, \Q'$) be a modified \ac{FBAS} computed from
        $(\V, \Q)$ by substituting every mention of $i$ in $\Q$ for $j$ and vice versa and
        additionally interchanging $\Q(i)$ and $\Q(j)$.
        If $\Q$ and $\Q'$ are equal, we define $i$ and $j$ as symmetric to one another as $\Q$ and
        $\Q'$ can only be equal if $\Q(i)$ and $\Q(j)$ are identical.
        In this case, the distribution function is \emph{symmetric} iff $\dist(i) = \dist(j)$.

        For example, in a symmetric \ac{FBAS} $(\V, \Q)$ where $\V = \set{0, 1,
        2}$ and
        $\forall k \in \V \colon \Q(k) = (\set {\V}, \emptyset,
        1)$ swapping nodes $0$ and $1$ results in $ \forall k \in \V \colon
        \Q'(k) = (\set {1, 0, 2}, \emptyset, 1)$.
        In this case, switching $\Q(0)$ and $\Q(1)$ has no effect as all $\Q(k)$ are the same.
        $\Q$ and $\Q'$ are equal; hence, $0$ and $1$ are also symmetric in $(\V,
        \Q)$.

    \item \label{req:dummy} \emph{Freeloader freeness}: Let $\Min{U_i}$ be the set of all minimal
        quorums in $(\V, \Q)$ that contains $i$.
        If $\Min{U_i} = \emptyset$ then $\dist(i)$ is \emph{freeloader-free} iff $\dist(i) = 0$.
        Clearly $i$ does not play a role in upholding the liveness of the
        \ac{FBAS}.
        Given that $i$ can leave and join the \ac{FBAS} without consequences for the \ac{FBAS},
        $\dist(i)$ should be $0$.
    \item \label{req:comp} \emph{Computational feasibility}: If $\dist(i)$ can be computed in
        polynomial time with respect to the number of nodes in the \ac{FBAS}, $\dist(i)$ is
        \emph{computationally feasible}.
    \item \label{req:correct} \emph{Correctness}: Assuming knowledge of $i$'s expected payoff $p$,
        then $\dist(i)$ is \emph{correct} iff $\dist(i) = p$.
        In anticipation of \cref{sec:design}, this is property is relevant when
        a distribution is approximated using probabilistic methods.
\end{enumerate}

\section{Fair Distribution of Rewards}%
\label{sec:design}

Having established a set of desirable characteristics of reward functions, we
now realize a reward distribution mechanism founded on \acf{CGT}.
We propose to measure a node's contribution to an \ac{FBAS} using the \emph{Shapley value} and
allocate rewards based on the computed values.
Although the Shapley value is one of multiple definitions of fairness, we choose
to use it as it is the only function that not only satisfies our symmetry and
freeloader freeness requirements, but also yields efficient and additive
distributions.
To this end, we recapitulate select formal notions surrounding cooperative games
and present our first contribution where we model an \ac{FBAS} as a cooperative game.
Subsequently, we define the reward distribution function
using the cooperative game model.
We also show that only a subset of nodes has non-zero Shapley values in the defined game.
This serves as an important observation as it enables an optimization for the
computation of Shapley values for larger \acp{FBAS}.

\subsection{Cooperative Game Theory}%
\label{sub:cgt}

Game theory models strategic interactions between \emph{players} as a
\emph{game}.
\ac{CGT} assumes that players benefit from cooperation as opposed to acting
individually and that the incentive to cooperate is great enough such that there
is no disadvantage in forming a single group consisting of all the
players~\cite{osborne2004introduction}.
A fundamental problem in \ac{CGT} is the \emph{fair} distribution of
\emph{collective payoffs} among the players.
To this end, let $\players = \set{1, 2,..., n}$ denote the set of players
with $n \geq 1$.

\begin{definition}[Coalition \cite{barron2013game}]\label{def:coaltion}
    A \emph{coalition} $\coalition$ is a subset $\coalition \subseteq \players$ of the set of
    players.
    The \emph{grand coalition} $\players$ is the coalition of all the players.  As there are $2^n$
    possible subsets of $\players$, there are also $2^n$ possible coalitions.
\end{definition}
As players are always assumed to be \emph{rational} (and not \emph{altruistic}),
a player will only join a coalition,
if the player's membership in the coalition is of benefit to the player.

\begin{definition}[Characteristic function of an $n$-player cooperative
    game \cite{barron2013game}]\label{def:char_func}
    Any function $\charfunc \colon 2^N \mapsto \mathbb{R}$ satisfying \\[1ex]
    \centerline
    {$\charfunc(\emptyset) = 0$ and $\charfunc(\players) \geq
    \sum\limits^{n}_{i=1}\charfunc(\set{i})$} \\[1ex]
    is a characteristic function of an
    $n$-player cooperative game.
\end{definition}
The characteristic function $\charfunc(\coalition)$ quantifies the coalition $\coalition$'s
\emph{worth} using a real-valued function.
The worth of the empty coalition is zero,
whereas the worth of the grand coalition $\players$ must be at
least equal to the sum of the individual payoffs.
This underpins the assumption that forming the grand coalition is not penalized
and may in fact be incentivized.

\begin{definition}[Cooperative Game with transferable utility,
    i.e., TU game \cite{osborne2004introduction}]\label{def:coop_game}
    A \emph{cooperative} game $G(\players, \charfunc)$ consists of a set of players $\players$ and
    a characteristic function $v \colon 2^N \mapsto \mathbb{R}$.
\end{definition}
As per \cref{def:char_func}, the characteristic function of a cooperative game defines the worth of
every coalition $\coalition$ that may form within the game.
A coalition's worth can be interpreted as the total \emph{payoff} or \emph{utility} that may be
freely distributed among the members of the coalition.
Such a game is said to have \emph{transferable payoff},
i.e., a coalition's payoff may be distributed
in multiple ways among the members of the coalition.

\begin{definition}[Simple Game \cite{barron2013game}]\label{def:simple_game}
    A \emph{simple game} is a cooperative game $G(\players, \charfunc)$ in which
    $\charfunc(\coalition) = 1$ or $\charfunc(\coalition) = 0$ for all coalitions in $G$.
    A coalition with $\charfunc(\coalition) = 1$ is called a \emph{winning coalition} and one with
    $\charfunc(\coalition) = 0$ is a \emph{losing coalition}.
\end{definition}
Simple games can, for instance, be used to model majority voting with $\charfunc(\coalition) = 1$ if
$\coalition$ has a majority of members, and $\charfunc(\coalition) = 0$ otherwise.

Since the characteristic function $\charfunc(\players)$ %
specifies the grand coalition's worth,
it is necessary to find some impartial method to divide this value among the
players signifying their contributions to the total worth.
Among other solution concepts~\cite{schmeidler1969nucleolus, neumann1944theory},
the \emph{Shapley Value}~\cite{shapley1951notes}
provides a solution for the set of $n$ players,%
\footnote{In the presence of smaller coalitions, the same solution concepts can
be applied analogously to the subgames defined by the coalitions.} while
recognizing that players may not be equally valuable to the coalition.
The Shapley value is each player's expected contribution to any possible
sequence of the $n$ players that may come together and form the grand coalition.
\begin{definition}[Shapley Value \cite{shapley1951notes}]\label{def:shapley}
    An allocation $\shapley(\charfunc) = (\shapley_1(\charfunc),...,\shapley_n(\charfunc))$ for a
    cooperative game $G(\players, \charfunc)$ is called the \emph{Shapley value} if
    \begin{center}
        $\shapley_i(\charfunc) = \sum\limits_{\set{\coalition \in \Pi^{i}}} [\charfunc(\coalition) -
        \charfunc(\coalition \setminus \set{i})] \cdot \cfrac{(|\coalition| - 1)!(|\players| -
        |\coalition|)!}{|\players|!},\ i = 1,2,...,|\players|$
    \end{center}
    where $\Pi^{i}$ is the set of all coalitions $\coalition \subseteq \players$ containing $i$ as a
    member.
\end{definition}
A player $i$'s Shapley value is equivalent to the average value $i$ adds to the
grand coalition when the grand coalition is formed in completely random order.

\subsection{FBAS as a Cooperative Game}%
\label{subsec:fbas_game}

Cooperation is imperative in the \ac{FBAS} system model: nodes make decisions about which other
nodes they need for consensus.
System-wide properties
such as liveness
are then a result of the sum of these individual
configurations.
We, therefore, argue that an \ac{FBAS} can be modeled as a cooperative game where players coalesce
into coalitions that establish and maintain desirable system properties.

We propose to model an \ac{FBAS} $(\V, \Q)$ as a cooperative game $G(\players, \charfunc)$ with a
characteristic function that uses the set of all quorums $\Quorums$ as measure of a coalition's
value.
For an \ac{FBAS}~$(\V, \Q)$, a cooperative game $G (\players, \charfunc)$ with player~$i \in
\players$, and an coalition~$\coalition \subseteq \players$, we use the following
notation:
\begin{equation*}%
\label{eq:cgt_notation}
\resizebox{1\columnwidth}{!}{$
  \begin{aligned}
    \Quorums \subseteq 2^{\V} & \colon\quad \text{the set of all quorums in}~(\V, \Q)~
      \text{as defined in \cref{def:quorum}} \\
    \charfunc(\coalition) & \colon\quad \text{the worth of the coalition}~\coalition \\
    \sspi_i & \colon\quad \text{player $i$'s Shapley-Shubik power index}\\
    \Pi^{i} \subseteq \Pset & \colon\quad \text{the set of all coalitions containing $i$ as a member}\\
    \W^{i} \subseteq \Pi^i  & \colon\quad \text{the set of all coalitions that win with player~$i$ and lose without $i$,}\\
      & \phantom{\colon}\quad
      \text{i.e.,} \coalition \in \W^i \iff \set{\coalition \in \Pi^i |
        \charfunc(\coalition) = 1, \charfunc(\coalition \setminus \set{i}) = 0}\\
  \end{aligned}$
}
\end{equation*}%

Given an arbitrary \ac{FBAS}, we map the set of nodes~$\V$
to the set of players~$\players$ directly, i.e., $\players = \V$.
In order to complete the definition of a cooperative game,
we need to provide a characteristic function~$v$
that quantifies every possible coalition~$\coalition$'s value,
i.e., $\charfunc \colon \Pset \to \mathbb{R}$.
We define a \emph{simple game} where a coalition $\coalition \subseteq
\players$ is a \emph{winning coalition} if it contains a quorum,
i.e., there is some quorum, $U\in\Quorums$ with $U \subseteq \coalition$.
$\coalition$ is a \emph{losing} coalition if it does not contain a quorum.
Formally, that is
\begin{equation}%
    \centering
    \label{eq:fbas_char_func}
    \begin{aligned}
        &\charfunc(\coalition) = \begin{cases}
            1 & \text{if } \exists U \in \Quorums \colon U
            \subseteq \coalition \\
            0 & \text{otherwise}
    \end{cases}
    \end{aligned}
\end{equation}

The definition of this characteristic function captures our main design goal:
The set of quorums $\Quorums$ contains the sets of nodes in the \ac{FBAS} that
play a role in upholding liveness of the system.
To this end, forming a coalition in this cooperative game is equivalent to nodes
defining quorum slices that result in a quorum in the corresponding \ac{FBAS}.
Given that an \ac{FBAS} already enjoys quorum intersection,
we reason that if a coalition $\coalition$ has no quorum,
it is of no worth to the \ac{FBAS}, %
hence $\charfunc(\coalition) =0$.
On the other hand, a coalition that has a quorum contributes to a common goal of
the \ac{FBAS}.

\subsection{Distribution of Rewards}%
\label{subsec:reward_dist}

We now define a function for the distribution of rewards based on the various
coalitions' worth in the game.
The computation of the distribution is a twofold procedure: we first determine each player's
\enquote{importance} using the Shapley-Shubik power index and then allocate rewards based on the
computed indices.

\subsubsection{The Shapley-Shubik Power Index}
\label{subsubsec:sspi}

Our cooperative game representation of an \ac{FBAS} is a simple game with a
superadditive characteristic function.
That means that the term $[\charfunc(\coalition) - \charfunc(\coalition \setminus \set{i})]$ of the
Shapley value function in \cref{def:shapley}
is always zero or one.
We can, therefore, simplify the Shapley value by
summing over only the coalitions where $[\charfunc(\coalition) - \charfunc(\coalition \setminus
\set{i})]$ equates to one.
These are coalitions $\coalition \in \W^i \subseteq \Pi^i$ that win with player
$i$ \emph{but} lose without player $i$.
Such a player is known as a \emph{critical} or \emph{pivotal} player as they
turn a losing coalition into a winning coalition when added to the coalition.
The remaining expression is better known as the \emph{Shapley-Shubik power
index}~\cite{shapley1954method} and defined for a player $i$ as follows.

\begin{definition}[The Shapley-Shubik Power
    Index \cite{barron2013game}]
    \label{def:sspi}\
    \begin{center}
        $\sspi_i = \sum\limits_{\set{\coalition \in \W^{i}}}
        \cfrac{(|\coalition| - 1)!(|\players| -|\coalition|)!}{|\players|!} \in
        \left[0, 1\right]$
    \end{center}
\end{definition}
\vspace*{1ex}

\subsubsection{Reward Distribution based on the Shapley-Shubik Power Index}%
\label{subsubsec:cgt_dist_func}

Given an \ac{FBAS} $(\V, \Q)$ in cooperative game form $(\players, \charfunc)$,
we calculate each player's power using the Shapley-Shubik power index and obtain
a real-valued index for each player $i$.
The set $\W^i$ for player $i$ is equivalent to
all subsets of $\V$ that
include node $i$ and only contain a quorum with node $i$'s membership.
The bigger the set of winning coalitions $\W^i$ for player $i$ is,
the higher their power index $\sspi_i$ will be,
and consequently, $i$'s share of the total reward.

We accordingly define a reward distribution function~$\dist(i)$
for an \ac{FBAS} $(\V, \Q)$ as the Shapley-Shubik
power indices of the corresponding cooperative game $(\players, \charfunc)$, i.e.,
\begin{equation}%
\label{eq:dist_func}
    \dist(i)  = \sspi_i
\end{equation}
The function $\dist(i)$ assigns a node $i$ its portion of a reward using its power index~$\sspi_i$.
A player~$i$ with $\sspi_i=1$ determines the outcome of the game
entirely on their own,
whereas $\sspi_i=0$ means that the player has no effect on the game's outcome.

\subsection{Relevant Nodes}%
\label{subsec:relevant_nodes}

We now show that only the set of \emph{top tier} nodes~$T$ of an \ac{FBAS}
can form winning coalitions for the game in \cref{eq:fbas_char_func}.
Recall that the top tier~$T \subseteq \V$ of an \ac{FBAS} ($\V, \Q$)
is defined as the union of all nodes in the set of all minimal quorums~$\Minquorums \subseteq 2^{\V}$, i.e., $T=\bigcup\Minquorums$ (cf. \cref{def:tt}).

\begin{theorem}[The set of critical players in $(\players, \charfunc)$ is equivalent to the top tier
    of the \ac{FBAS}$(\V, \Q)$]%
    \label{theorem:tt}
    Let $T$ be the top tier of an \ac{FBAS} $(\V, \Q)$ and $(\players, \charfunc)$ its
    cooperative game form.
    Let $\Minquorums \subseteq 2^{\V}$ be the set of minimal quorums for FBAS $(\V, \Q)$ as defined
    in \cref{def:minquorums} and $\Quorums \subseteq 2^{\V}$ the set of quorums.
    The set $\Pi^{i} \subseteq 2^{\players}$ contains all the coalitions that include player $i$ and
    $\W^i \subseteq \Pi^{i}$ all coalitions for which player $i$ is critical.
    Then $\forall i \in \V: i \notin T \iff \W^i = \emptyset$.

    \begin{proof}~
      \begin{description}[labelsep=2pt]
        \item [$\Rightarrow \colon$] Assume $\W^i \neq \emptyset$.
          As per the definition of $\W^i$ in \cref{eq:fbas_char_func}, every coalition
          $\coalition \in \W^i$ contains a quorum and is hence a \emph{winning} coalition.
          Furthermore, we know that $i$ is a critical player in every such coalition,
          i.e., $\forall \coalition \in \W^i \colon \charfunc(\coalition \setminus \set{i}) = 0$.
          A winning coalition $\coalition \in \W^i$ must therefore contain some minimal quorum
          $\Min{U} \in \Minquorums \colon i \in \Min{U}$ for if this were not the case,
          $\Min{U} \setminus \set{i}$ would still be a quorum and $\charfunc(\coalition
          \setminus \set{i}) = 1$.
          Since $i \in \Min{U}$, it follows that $i \in T$ which is a contradiction.

        \item [$\Leftarrow \colon$] Since $W^i = \emptyset$ there is no set that fulfills
          \cref{eq:fbas_char_func} only because of player $i$'s membership, i.e.,
          $\forall \coalition \subseteq \Pi^{i} \colon \charfunc(\coalition) = 1 \land
          \charfunc(\coalition \setminus \set{i}) = 1 \lor \charfunc(\coalition) = 0 \land
          \charfunc(\coalition \setminus \set{i}) = 0$.
          Hence, there is no set $U_1 \subseteq 2^{\V}$ that fulfills $U_1 \cup
          \set{i} \in \Quorums$ \emph{while} $U_1 \setminus \set{i}
          \notin \Quorums$.
          If player $i$ is critical for some quorum, then $i$ is also critical for some
          minimal quorum because a minimal quorum is a quorum in which each of its members is
          critical.
          Therefore, there cannot be any minimal quorum $\Min{U_1}$ such that $i \in \Min{U_1}$.
          It follows that if $\W^i = \emptyset \implies i \notin T$.
        \end{description}
    \end{proof}
\end{theorem}
It is, therefore, sufficient to initialize the cooperative game of an \ac{FBAS} with $\players = T$.
The cooperative game $(\players, \charfunc)$ as defined in \cref{eq:fbas_char_func} of \ac{FBAS}
($\V, \Q$) with top tier $T$ is equivalent to the cooperative game $(T, \charfunc)$.
All nodes in $\V \setminus T$ will always have a power index of zero as they have no winning
coalitions.

\subsection{Computing the Shapley-Shubik Power Index}%
\label{subsec:impl}

In order to calculate the Shapley-Shubik power index as defined,
we can use an \emph{enumeration} algorithm:
This algorithm enumerates all $\Pset$ possible coalitions for a game $(\players,\charfunc)$
and determines if a given coalition is \emph{winning}
according to the characteristic function defined in \cref{eq:fbas_char_func}.
Such enumeration algorithms, however, have an exponential time complexity
of~$\mathcal{O}(2^{|\players|})$, rendering the algorithms unusable as the
number of players increases.
We therefore cannot expect to compute the power indices
efficiently \emph{and} accurately.

In order to tackle the efficiency problem with the enumeration algorithm,
we propose to calculate the Shapley-Shubik power index based
on the random sampling algorithm by Castro et al.~\cite{castro2009polynomial}.
The algorithm approximates the Shapley values\footnote{Recall that the Shapley-Shubik power index is a
simplified form of the Shapley value in the case of simple games (see \cref{subsubsec:sspi}).}
using $m$ randomly drawn permutations of the grand coalition.
Unlike enumeration algorithms, the sampling algorithm approximates players'
Shapley values in polynomial time subject to the efficient computation of the
game's characteristic function.
The proposed characteristic function in \cref{eq:fbas_char_func} essentially
consists of checking that a given coalition contains a quorum which can be done
in polynomial time.
We refer interested readers to works such as~\cite{florian2022sum,
gaul2019mathematical, lachowski2019complexity} for results
pertaining to the computational complexity of problems related to the properties
of \acp{FBAS}.

Although the sampling algorithm provides asymptotic error bounds,
Maleki et al. provide finite error bounds for the above sampling algorithm when
either the variance or range of marginal contributions is
known~\cite{maleki2013bounding}.
They also show that approximations can be improved using
\emph{stratified} sampling when the set of permutations can be divided into
strata with comparable marginal gains.
In practice, all top tier nodes tend to be configured
identically~\cite{florian2022sum, ndolo2021crawling} such that their marginal
contributions do not differ.
We therefore do not expect significant gains from stratified sampling methods.

\subsection{Discussion}%
\label{subsec:discussion}

In the following, we address and discuss limitations of the proposed
distribution function.
A malicious node that is part of the top tier is able to manipulate its own
Shapley value by altering the set of minimal quorums using Sybil nodes.
A malicious node $k$ can create a set of $j \geq 1$ Sybil nodes
each configured with a single quorum slice $\set{i, k}, i \in \set{1,...,j}$.
This results in a higher Shapley value for the malicious node $k$ and
consequently, a bigger share of the reward.
The Sybil nodes' Shapley values will be zero as they are not critical for any
coalition.
An adversarial node outside the top tier is not able to manipulate the
distribution function as the formation of new minimal quorums is not possible
without the cooperation of the existing nodes' while maintaining
safety~\cite{florian2022sum}.

We consider the described possibility to be inherent to the \ac{FBAS} model as
there is nothing that prohibits a top tier node from constructing quorum slices
with Sybil nodes.
In fact, we conjecture that it is generally not possible to distinguish between
Sybil and non-Sybil nodes in such cases.
This limitation would therefore apply to any reward distribution approach.
While extending an FBAS-based system with monetary incentives may provide
additional motivation to mount such attacks, it represents an interesting option
towards improving the resilience of some FBAS-based systems.

The practical relevance of such attacks might be limited.
While a Sybil attack can be mounted by any node in the FBAS, it is only
beneficial to a node that is part of the top tier
(cf.~\cref{subsec:relevant_nodes}).
In practice, the top tier of these systems is a fairly fixed set of
nodes that forms over time with new nodes able to join the top tier only with
the cooperation of existing members (see~\cite{florian2022sum} for
proofs and discussions on the evolution of a top tier).
Node operators in the Stellar and MobileCoin networks exercise off-chain
coordination of aspects pertaining to configurations and monitoring.
This means that a (malicious) node is not simply able to infiltrate the top tier
of such an FBAS.
If the malicious node is already a member of the top tier and has been
identified as such, an attack can be hampered by reconfiguration of the other
top tier nodes facilitating the removal of the attacker from the top tier.

Determining and quantifying the extent to which specific FBAS instances are
vulnerable to reward-inflating Sybil attacks is an interesting research question
for future work.

\section{Evaluation}%
\label{sec:evaluation}

In the following, we evaluate the presented reward function
against the set of requirements defined in \cref{subsec:requirements}.
We evaluate symmetry and freeloader freeness formally,
and computational feasibility and correctness empirically.
To this end, we implemented the proposed reward scheme and provide
a command-line framework
that is available on Github.\footnote{\fbasdistlink}
We generate two types of synthetic \acp{FBAS} with an increasing number of nodes
consisting solely of a top tier as was done in~\cite{florian2022sum} and define
a node $i$'s quorum function using a tuple notation reminiscent of a quorum set
(cf. \cref{subsec:preliminaries}).
The first \ac{FBAS} is similar to traditional $3f+1$ quorum systems
where all nodes have identical quorum sets tolerating the same number of
failures:
\begin{align*}
  \forall i \in \V \colon \Q(i) = \left(\V, \emptyset, \optithresh{\V}\right)
\end{align*}
This structure is currently found in MobileCoin's public network~\cite{ndolo2021crawling}.
The second kind of \ac{FBAS} approximates the structure of the Stellar network's
top tier in which each node is operated by an organization.
Furthermore, each organization typically operates three physical nodes
configured with $2f+1$ crash-tolerant inner quorum sets:
\begin{align*}
    &\V = \set{i_0, i_1,...,i_{n-1}}, n = 3m \\
    &\I = \set{(\set{i_{3j}, i_{3j+1}, i_{3j+2}}, \emptyset, 2) \mid i \in [0,
    m)}\\
    &\forall i \in \V \colon \Q(i) = \left(\emptyset, \I,
        \optithreshraw{m}\right)
\end{align*}

\subsection{Symmetry}%
\label{sub:symmetry}

As per Requirement~\ref{req:eq}, the reward distribution is \emph{symmetric} if
$\dist(i)$ = $\dist(j)$ for two symmetrical nodes $i, j \in \V$ in \ac{FBAS} $(\V, \Q)$.
The distributions $\dist(i)$ and $\dist(j)$ are equal if
the Shapley-Shubik power indices $\sspi_i$ and $\sspi_j$ are equal as well.
Therefore, if $\sspi_i = \sspi_i \implies \dist(i) = \dist(j)$.
We show that if $i, j \in \V$ are symmetric to each other, then $\sspi_i = \sspi_j$.
    \begin{proof}
        Let $i, j \in \V$ be symmetric to each other in \ac{FBAS} $(\V, \Q)$.
        Because $\Q = \Q'$ after the substitution $i, j$, it can be shown that the respective set of
        all quorums $\Quorums$ and $\Quorums'$ are equal.
        Because $\Quorums = \Quorums'$, then $i$ and $j$ are members of the same number of quorums
        with the same cardinality.
        It follows that the sets of coalitions $\W^i \subseteq \Pi^i \subseteq 2^{\V}$ and $\W^j
        \subseteq \Pi^j \subseteq 2^{\V}$ for which $i$ and $j$ are critical, must also be
        equivalent. %
        Hence, $\sspi_i$ is inevitably equal to $\sspi_j$.
        This proves that, for two nodes $i, j$ symmetric to one another, $\sspi_i(i) = \sspi_j(j)$.
    \end{proof}
A node's allocation using the function in \cref{eq:dist_func} is its power index.
Hence, if $\sspi_i = \sspi_j$ then $\dist(i)$ and $\dist(j)$ are equal.
This distribution function is therefore \emph{symmetric}.

\begin{figure*}[!t]
    \begin{minipage}[t]{\columnwidth}
        \includegraphics[]{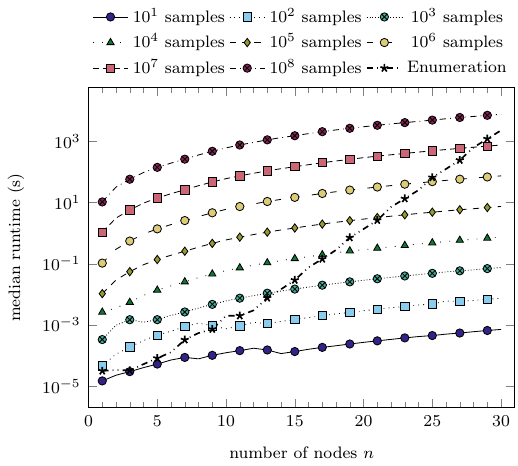}
        \caption{Runtime for FBASs resembling classical $3f+1$ quorum systems.}%
    \label{fig:performance_mobilecoin}
    \end{minipage}
    \begin{minipage}[t]{\columnwidth}
        \includegraphics[]{performance_results_mobilecoin.pdf}
        \caption{Runtime for FBASs similar to the top tier of the Stellar public
    network.}%
    \label{fig:performance_stellar}
    \end{minipage}
\end{figure*}

\subsection{Freeloader Freeness}%
\label{sub:freeloader}

We characterize a node $i \in \V$ as a \emph{freeloader}
if the set of minimal quorums $\Min{U_i} \subseteq 2^{\V}$
containing $i$ is empty as per Requirement~\ref{req:dummy}.
For an \ac{FBAS} $(\V, \Q)$, a freeloader is a player in the equivalent game $(\players,
\charfunc)$ with a non-zero power index despite not being critical for any coalition.

From \cref{theorem:tt}, it follows that the set of top tier nodes $T \subseteq \V$ is equivalent to
the set of critical players.
All other nodes in $\players \setminus T$ are not decisive in any coalitions and have Shapley-Shubik
power indices equal to zero.
It follows directly that our reward distribution function is
\emph{freeloader-free}.

\subsection{Computational Feasibility}%
\label{sub:computation}

We examine the \emph{computational feasibility}
using our implementation and perform runtime tests.
We generate synthetic \acp{FBAS} with up to $n \leq 30$ nodes and
for each corresponding game $G$ via direct enumeration and approximation.
We draw $m = 10^k, k \in [1..8]$, samples of the grand coalition for the approximation algorithm.
We compute distributions for all players in each $G$ with $n$ players and
measure the runtime using the operating system clock on our Linux-based
evaluation environment.
We repeat the experiments $10$ times for each $G$
and calculate the median runtime.
All computations were carried out single-threaded
on server-class hardware.

The results of our measurements are shown in \cref{fig:performance_mobilecoin}.
and \cref{fig:performance_stellar}.
As expected, the computation time of the power indices via enumeration
increases exponentially with the number of nodes in the \acp{FBAS}.
At the time of writing, the real-world networks MobileCoin and Stellar
comprise a top tier of $10$ and $23$ nodes respectively.
The computation of the power index via enumeration would therefore require $\leq
1s$ and $\approx 56s$, respectively.
We consider the runtime of the enumeration algorithm to be acceptable for
currently deployed network sizes,
but not feasible if we expect larger \acp{FBAS}. %

We observe a more favorable trend for an increasing number of nodes with the
approximation algorithm as was expected.
Nonetheless, the computation time depends on the sample size:
We observe that increasing the sample size by an order of magnitude
slows the computation down by a factor of $\approx 10$.
In particular for larger networks, we consider the approximation of the power
indices to be computationally feasible.
In anticipation of our correctness evaluation in \cref{sub:correctness}, the
approximation algorithm strikes a satisfactory balance between runtime and error
rate for sample sizes between $10^4$ and $10^6$.

\subsection{Correctness}%
\label{sub:correctness}

As per Requirement~\ref{req:correct}, a distribution function~$\dist(i)$ is defined
as \emph{correct}, if $\dist(i) = p$ assuming node $i$'s expected payoff $p$ is known.
Naturally, this property is only relevant to the approximation algorithm.
In order to evaluate the accuracy of the approximation, we follow an empirical
approach running a series of experiments and compare the approximated Shapley
values with the exact values.
We are again interested in the impact of the sample size~$\samples$
and the number of nodes~$n$.

For every cooperative game $G$ of a synthetic \ac{FBAS} with $n \leq 30$ nodes,
we proceed with our experiments as follows:
We calculate the exact Shapley-Shubik values $\sspi_i(\charfunc)$
as well as the approximated values $\hat\shapley_i(\charfunc)$
for every player~$i$ in~$G$ using~$\samples \in 10^k$, $k \in [1..8]$, samples.
We then calculate each player $i$'s \ac{MPE} for $G$ as follows
\begin{align*}
    \text{\ac{MPE}} =\frac{1}{n}\sum\limits^{n-1}_{i=0}\frac{\mid
    \hat\shapley_i(\charfunc) - \sspi_i(\charfunc) \mid}{\mid\sspi_i(\charfunc) \mid}
\end{align*}
We repeat the experiments $50$ times for $G$
and calculate the \ac{MMPE} as in~\cite{campen2018new}.

\begin{figure*}[!t]
    \begin{minipage}[t]{\columnwidth}
    \includegraphics[]{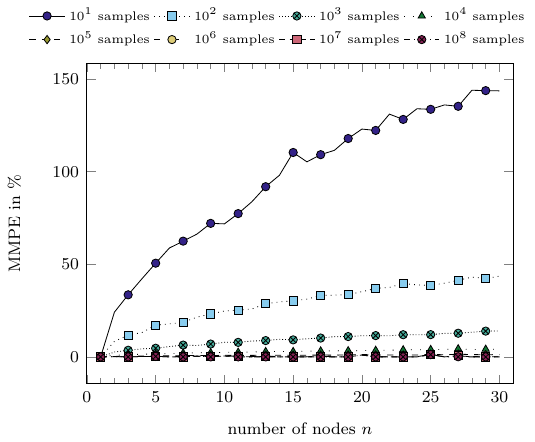}
    \caption{The \acp{MMPE} for FBASs resembling classical $3f+1$ quorum systems.}%
    \label{fig:error_mobilecoin}
    \end{minipage}
    \begin{minipage}[t]{\columnwidth}
        \includegraphics[]{error_results_mobilecoin.pdf}
        \caption{The \acp{MMPE} for FBASs similar to the top tier of the Stellar
        network.}%
        \label{fig:error_stellar}
    \end{minipage}
\end{figure*}%

\cref{fig:error_mobilecoin} and \cref{fig:error_stellar} depict the results of
our experiments.
We abstain from adding error bars to the plots for the sake of readability.
As expected, the results show that the relative errors are especially high
when approximations are performed with a small number of samples.
The number of permutations $n!$ grows rapidly as we increase $n$, and thus, a a
constant $\samples$ becomes an even smaller fraction of the total population of
permutations.
However, the errors increase relatively slowly compared to the number of nodes.
For example, approximations
with $5$ and $20$ nodes performed with $10^3$ samples have \acp{MMPE} of
$\approx 4.7\%$ and $\approx 10.5\%$ respectively.
This means that a factor $4$ increase in the number of nodes only leads to a factor $\sim2$ increase
in the \ac{MMPE}.
Starting from $\samples = 10^5$,
the \acp{MMPE} remain consistently below $1.6\%$ for all tested~$n$.

We conclude that for a constant $n$, the accuracy of the approximation increases
as larger samples are used.
For a constant number of samples, the error of the approximations rises as $n$
rises.
While approximations conducted with $m \geq 10^7$ samples have a near-zero error
regardless of $n$, the accuracy comes at the cost of runtime.
We therefore regard approximations performed with $m \geq 10^5$ samples as correct but emphasize the
importance of considering the runtime when choosing the sample size.
Sample sizes between $10^4$ and $10^6$ offer a reasonable compromise between
correctness and runtime and allow some leeway in either of the parameters.

\section{Related Work}%
\label{sec:related_work}

Blockchain-based \ac{P2P} networks typically offer a block reward
as an \emph{incentive} to miners and validators
in exchange for the resources they utilize to create new blocks.
Incentivization schemes have, therefore, been studied by several scholars,
including models for specific consensus mechanisms
such as \acf{PoW} and \acf{PoS}~\cite{ketsdever2019incentives, huang2021do}
and broader discussions of incentivization in \ac{P2P} data
networks~\cite{daniel2021ipfs}.
Multiple works study incentivization in collaborative schemes such as \ac{PoW}
mining pools~\cite{romiti2019deep, rosenfeld2011analysis} and \ac{PoS} stake
pools~\cite{bruenjes2020reward}.
To the best of our knowledge, no explicit incentivization scheme for the FBAS setting
has been proposed so far.

For FBAS-based systems, Kim et al.~\cite{kim2019stellar} hypothesize
that reward schemes could attract more validators,
which, in turn, increases the security %
by promoting decentralization and fault tolerance.
As an opposing view to the introduction of explicit rewards, Ketsdever et
al.~\cite{ketsdever2019incentives} and Babaioff et al.~\cite{babaioff2011on}
argue that monetary rewards may reintroduce some of the problems they initially
try to solve, e.g., centralization.
We consider this discussion to be out of the scope of the presented work,
focusing on the investigation of to what extent a fair rewards distribution is
at all possible in FBAS-based systems.

While no explicit efforts have been devoted to incentivization in \ac{FBAS}-based systems,
a few works have dealt with identifying pivotal nodes based on node
centrality~\cite{kim2019stellar, gaul2019mathematical}.
Although it is possible to adapt such heuristic ranking approaches to bootstrap
a reward distribution, we focus on developing a reward distribution approach
founded on game theory to provide rigid fairness guarantees and manipulation
resistance.
Methodically related to our approach,
Bracciali et al.~\cite{bracciali2021decentralization} investigate
the decentrality of open quorum systems.
They apply the Penrose-Banzahf power index and
present bounds on the decentrality in open quorum systems.
Bracciali et al. are, however, interested in a fundamentally different question
than we are in this paper.
Moreover, we define a more general, parameterless model for generic \ac{FBAS}-based systems
and use a different concept to measure power.

\section{Conclusion}%
\label{sec:conclusion}

We formulated a reward distribution function for the \acp{FBAS} model.
We drew on methods from \acl{CGT} and defined a cooperative game representation of an \ac{FBAS}
measuring contributions to the attainment of system-wide properties.
In contrast to heuristic ranking approaches, our presented model offers a means to objectively
and precisely quantify a node's power in an \ac{FBAS}.
By defining a function for the dissemination of rewards, we demonstrated that the
\ac{FBAS} model is compatible with incentives.

\IEEEtriggeratref{16}
\bibliography{paper.bib}

\end{document}